\documentclass[twocolumn]{revtex4-1}
\usepackage{graphicx}
\begin{document}
\title{Non-thermal $p/\pi$ ratio at LHC as a consequence of hadronic final state interactions}
\author{Jan Steinheimer}
\email{jsfroschauer@lbl.gov}
\affiliation{Lawrence Berkeley National Laboratory, 1 Cyclotron Road, Berkeley, CA 94720, USA}
\affiliation{Frankfurt Institute for Advanced Studies, Johann Wolfgang Goethe-Universit\"at, Ruth-Moufang-Strasse 1, 60438 Frankfurt am Main}
\author{J\"org Aichelin}
\affiliation{SUBATECH, UMR 6457, Universit\'{e} de Nantes, 
Ecole des Mines de Nantes, IN2P3/CNRS. 4 rue Alfred Kastler, 
44307 Nantes cedex 3, France}
\affiliation{Frankfurt Institute for Advanced Studies, Johann Wolfgang Goethe-Universit\"at, Ruth-Moufang-Strasse 1, 60438 Frankfurt am Main}
\author{Marcus Bleicher}
\affiliation{Institut f\"ur Theoretische Physik, Johann Wolfgang Goethe-Universit\"at, Max-von-Laue-Strasse 1, 60438 Frankfurt am Main}
\affiliation{Frankfurt Institute for Advanced Studies, Johann Wolfgang Goethe-Universit\"at, Ruth-Moufang-Strasse 1, 60438 Frankfurt am Main}

\begin{abstract}
Recent LHC data on Pb+Pb reactions at $\sqrt s_{NN}=2.7$~TeV suggests that the $p/\pi$ is incompatible with thermal models. We explore several hadron ratios (K/$\pi$, $p/\pi$, $\Lambda/\pi$, $\Xi/\pi$) within a hydrodynamic model with hadronic after burner, namely UrQMD 3.3, and show that the deviations can be understood as a final state effect. We propose the $p/\pi$ as an observable sensitive on whether final state interactions take place or not. The measured values of the hadron ratios do then allow to gauge the transition energy density from hydrodynamics to the Boltzmann description. We find that the data can be explained with transition energy densities of $840 \pm 150$~MeV/fm$^3$.
\end{abstract}
\maketitle
\section{Introduction}
With the start of the LHC physics program three years ago, the field of high energy nuclear physics, and especially heavy ion physics, has gone into a new era. It is now possible to explore the properties of Quantum-Chromo-Dynamics (QCD) at unprecedented particle densities and temperatures. The three large scale experiments ALICE, ATLAS and CMS have provided novel data on the transverse expansion dynamics, e.g. elliptic flow, jet attenuation far beyond the reach of RHIC, high quality charm and bottom data and the suppression of quarkonia \cite{Aamodt:2010cz,Abelev:2012rv,Chatrchyan:2011sx,Chatrchyan:2011pe}. Many of these data could be understood in terms of pQCD inspired models, parton cascades and hydrodynamic approaches \cite{Hirano:2010je,Schenke:2011tv,Petersen:2011sb,Uphoff:2010sh,Werner:2010aa,Konchakovski:2012yg}. 

A long standing alternative approach to interpret bulk hadron multiplicities is based on the statistical or thermal model of hadron yields \cite{Andronic:2008gm,Andronic:2008gu,Andronic:2011yq,Becattini:2000jw,Cleymans:1996cd}. Here one assumes the production of a thermalized fireball with a common temperature, $T$, a common baryo-chemical potential $\mu_B$, and a fixed volume. Under these assumptions one can extract the fireball parameters $T$ and $\mu_B$ from the measured hadron ratios. At energies below the current LHC energy a very large number of measured hadron ratios is compatible with the assumption that hadrons are produced from such a thermalized source. 
Deviations from the thermal predictions, as seen for resonances, could be interpreted as due to rescattering of one of the decay products not allowing for an experimental reconstruction of the resonances. 

At LHC energies this is different. Here, two of the elementary ratios, the $\overline{p}/\pi$ and $p/\pi$-ratios, show clear deviations from the predictions of the thermal model \cite{Andronic:2011yq,sqm} whose parameters are fitted to other particle ratios. It has been proposed that this deviation can be due to different hadronization temperatures of light and strange quarks \cite{Ratti:2011au} (however such an effect should be visible also at lower energies).

In this letter we suggest that this deviation is due to out of equilibrium final state interactions of the hadrons after the break-up of the thermal fireball. Discussions on hadronic final state interactions are ongoing since hadronic afterburners where first introduced. It is the primary aim of our paper to show the way how this question can be settled by a detailed analysis of the experimental data, i.e. the $p/\pi$ ratio. Final state interactions are, at lower energies, too weak to largely distort this ratio, though a recent investigation within the statistical model suggests that hadronic final state interactions should not be neglected for anti-particle yields even at SPS and RHIC energies \cite{Becattini:2012sq}. We will furthermore show that this
ratio allows for the experimental determination of one of the key quantities in present day transport theories at this energy:
The energy density at which the systems ceases to be able to maintain a local equilibrium and passes to the phase of
non-equilibrium final state interactions. This energy density is called freeze out or transition energy density. In the hydrodynamical approaches this is the energy density where hadrons are formed. We elucidate this idea by employing a combined hydrodynamic plus hadronic cascade model.

\section{UrQMD hybrid approach}
For the present calculations we employ the UrQMD 3.3 hybrid model to central Pb+Pb reactions at the current LHC energy of $\sqrt s=2.75$~TeV. The UrQMD hybrid approach extends previous ansatzes to combine hydrodynamic and transport models for relativistic energies by combining these approaches into one single framework for a consistent description of the dynamics \cite{Paiva:1996nv,Aguiar:2001ac,Teaney:2001av,Socolowski:2004hw,Hirano:2005xf,Hirano:2007ei,Bass:1999tu,Nonaka:2006yn,Dumitru:1999sf,Werner:2010aa}. The simulations starts from a PYTHIA generated initial condition \cite{Bass:1999tu,Dumitru:1999sf} which is mapped onto a hydrodynamic energy momentum tensor assuming local equilibration. In the next step the ideal hydrodynamic equations are solved \cite{Rischke:1995ir} until a local transition criterion is reached. The transition criterion from the hydrodynamic description to the transport stage is defined by the local energy density in the rest frame of the cells, i.e. $\epsilon_{\rm transition}=n*\epsilon_0$, with $\epsilon_0$ being the groundstate energy density, the parameter $n$ has usually been set to a value of $n=5$. By Monte-Carlo sampling of the Cooper-Frye distribution, the local densities are than transformed into hadrons and evolved in the hadronic cascade until all reactions cease. For a detailed description of the model and how it can be extended to LHC energies we refer to \cite{Petersen:2008dd,Steinheimer:2009nn,Petersen:2011sb,Petersen:2011fp,Petersen:2012qc}

\section{Effect of the cascade stage}
\begin{figure}[t]
\includegraphics[width=0.5\textwidth]{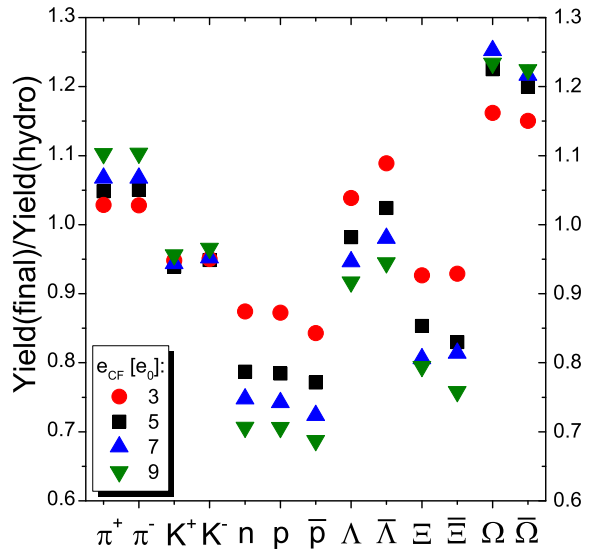}
\caption{\label{absorb}(Color online) Ratios of particle multiplicities in the final state scaled by the particle multiplicity directly after the hydrodynamic stage for central Pb+Pb reactions at $\sqrt s=2.75$~TeV. The different symbols denote different transition energy densities in units of $\epsilon_0$.}
\end{figure}

Figure \ref{absorb} shows the ratios of particle multiplicities in the final state scaled by the particle multiplicity directly after the hydrodynamic stage for central Pb+Pb reactions at $\sqrt s=2.75$~TeV. In both cases the hadron yields include the feed down from all resonances in the model. The different symbols denote different transition energy densities in units of $\epsilon_0 = 145 MeV/fm^3$. One observes a dependence of the yields on the transition energy density. For early transitions, i.e. at a higher energy density, the modifications are stronger than for a later transition, i.e. a lower energy density. While the meson yields are only modified on a 10\% level, for transition energy densities between $3-9\epsilon_0$, a stronger effect can be observed for (anti-)protons. $\Lambda$'s on the other hand are not as much modified. This indicates, that the dominant process in the decoupling phase, following the Cooper-Frye transition, is the annihilation process, reflected in the strong dependence of the proton final yield on the transition density. It is expected that in an expansion the annihilation processes are the first which signal that the systems leaves chemical equilibrium \cite{Rapp:2000gy,Rapp:2001bb,Greiner:2002nc,Rapp:2002fc}. Usually a baryon-anti baryon annihilation creates a couple of pions ($n \approx 5$) \cite{Blumel:1994yj,Dover:1992vj} and therefore the inverse process is highly density dependent and a high density is required to maintain chemical equilibrium \cite{Cassing:2001ds}. This is in contrast to resonances which dominantly disintegrate into two particles and therefore can maintain their equilibrium distribution for a longer time. The small change in the $\Lambda$ yield can therefore be explained by regeneration through resonance excitations in the final state, a process that is not allowed for protons which have to be created as a $p$-$\overline{p}$ pair. With this, our results agree with previous calculations, where the UrQMD model was used for the final state of a hydrodynamical calculation \cite{Bass:2000ib}.\\

In previous studies \cite{Rapp:2000gy,Rapp:2001bb,Greiner:2002nc,Rapp:2002fc} the importance of the multi meson fusion reaction for describing anti-baryon yields at the SPS (and lower) energies was pointed out. At those energies the net baryon number is still considerably large and the number of protons larger than that of the anti-protons. The pion density at hadronization, however, should not change considerably from the SPS to LHC as the transition temperature $T_c$ increases  only very slowly with beam energy. Because the inverse reaction $5 \pi \rightarrow p + \overline{p}$ is proportional to the fifth power of the pion density, we can expect the relative contribution from multi pion scattering to the anti-proton yield to decrease with beam energy, as there are considerably more anti-protons produced, at hadronization, at the LHC than at the SPS.\\
In the following we want to clarify the differences and the apparent conflict of our work with that of Rapp and Shuryak \cite{Rapp:2000gy,Rapp:2001bb,Rapp:2002fc}. The relevance of the annihilation and its back reaction is determined by the dynamical evolution of the phase space densities of the involved particles. In the work of Rapp and Shuryak this evolution is simplified to a certain degree, and therefore, already the annihilation rate is significantly overestimated. A previous investigation with our model showed that anti-proton absorption, at top SPS energies, can very well be described and does not lead to such a drastic depletion of anti-protons as predicted by Rapp and Shuryak \cite{Becattini:2012sq}. Consequently, due to the overestimated annihilation, the pion phase space in their work turns out over saturated. As the back reaction is proportional to the fifth power of the pion density it is also largely overestimated to compensate the too large annihilation rate. The final value for the anti-proton yield, obtained by Rapp and Shuryak, turns out to be not much different from the results with our model, presented in \cite{Becattini:2012sq}. 
Considering that the SPS data, referred to in \cite{Rapp:2000gy}, have been corrected since, to accommodate weak decays, by roughly a factor of 0.5, the back reaction does not play such a significant role as previously proposed.
 
To give a quantitative bound on the systematic error of our result, due to lack of detailed balance, we investigated the proton loss in a static system, within UrQMD. For this we initialized systems in chemical equilibrium, corresponding to the average densities/temperatures at our different transition criteria. We then let the system evolve in a static box, i.e. without expansion, and observe a depletion of the proton yield in that box (without resonance decays), due to lack of detailed balance. We can observe a loss of protons proportional to the time which is largest at the highest temperature, as expected. From these results we extract the temperature dependent proton loss due to lack of detailed balance $dN_p/dt (T)$. Because the system created at the LHC is fast expanding it remains at the highest temperature only for a short time. An estimate on the time evolution of the density/temperature can be obtained from extending the hydro simulation after the transition. We use the time dependence of the temperature, down to 130 MeV, to estimate the upper bound on the total number of protons lost due to lack of detailed balance by integrating $dN_p/dt (T)$ over the evolution of the system. We find it to be around $8\%$ of the starting value (for the highest transition temperature), considered that $50\%$ of all protons come from resonance decays. From this we conclude that our results, presented in figure \ref{absorb}, are qualitatively robust and quantitatively off by less than 8\%.\\
The contribution from the backward reaction should even be smaller, considering that the system is not static but fast expanding with a large collective velocity which drives the particle freeze out and should suppress the multi pion reactions. This has been shown in \cite{Rapp:2001bb} where the chemical relaxation times for the anti-protons quickly exceeded the fireballs lifetime because multi-pion annihilation is not frequent enough to counteract the $p+\overline{p}$ annihilation.\\

\begin{figure}[t]
\includegraphics[width=0.5\textwidth]{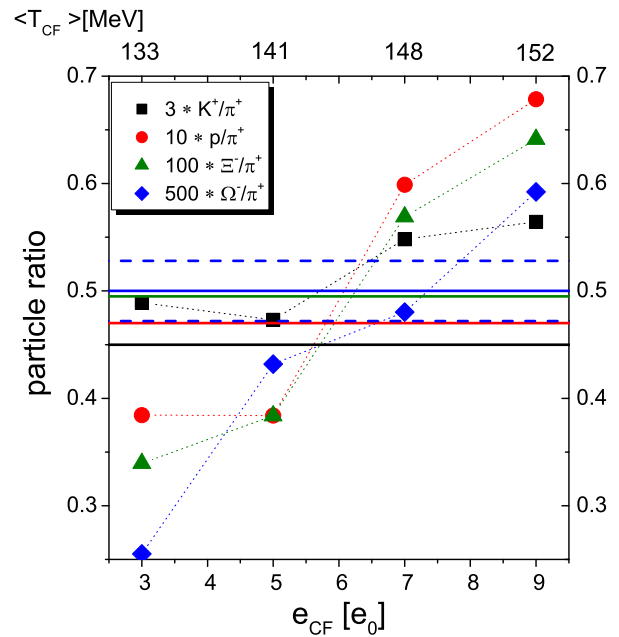}
\caption{\label{ratio}(Color online) Hadron ratios as a function of the transition energy density for central Pb+Pb reactions at $\sqrt s=2.75$~TeV. The symbols denote the calculations, the line depict the experimental data.}
\end{figure}

Figure \ref{ratio} shows the final hadron ratios (scaled for visibility) as a function of the transition energy density for central Pb+Pb reactions at $\sqrt s=2.75$~TeV. The average production temperature $\left\langle T_{CF} \right\rangle$, corresponding to the transition energy density boundary is shown on the upper axis of figure \ref{ratio}. The symbols denote the calculations, the solid lines depict the experimental data (the dashed lines correspond to the experimental errors in the $\Omega/\pi$) \cite{sqm}. Depending on the transition energy density, i.e. on the lifetime of the hadronic stage, the baryon/meson  ratios are systematically modified. The higher the transition energy density  the higher the density and therefore the more probable that the baryons find an annihilation partner.
In a rapidly expanding system the inverse reaction becomes rare and therefore the deficit with respect to thermal model prediction remains. The lower the transition energy density the lower is the density  and the shorter is the life time of the hadronic stage. Therefore it is more difficult for the baryons to find an annihilation partner and  the ratios approach 
the values expected from a thermal distribution at the transition energy density.\\
All ratios approach the experimental values if the transition energy density is around  $\epsilon=6\epsilon_0$,
close to our standard value ($\epsilon=5\epsilon_0$).\\
From this investigations, we conclude that the apparent inconsistency of experimental  $p/\pi$-ratio with the thermal model predictions can be explained by the hadronic final state interaction, which modifies the hadron yields. Even more, their values as compared to the thermal model predictions allows for an experimental determination of the transition energy density. By comparison to the experimental data we can fix the transition energy density in the UrQMD hybrid model to $840 \pm 150$~MeV/fm$^3$.

\section{Summary}
The apparent non-thermal $p/\pi$ ratio observed at LHC has been investigated. We found that the $p/\pi$- ratio is strongly modified due to the late stage hadronic effects. In this stage the hadrons fall out of equilibrium until they finally decouple.
One can summarize the main points of our findings as:
\begin{itemize}
\item The description of the LHC data requires the inclusion of an out-of-equilibrium final hadronic stage.
\item Signals on where this transition takes place are not washed out completely during this out-of-equilibrium 
final hadronic stage.
\item This signal can and has to be used to calibrate the different models. It determines where in a given model (and if
consistency between different models can be achieved in anture) the
transition between the eq. and the out-of-eq. phase takes place.
\end{itemize}

Even more important our findings impose a benchmark to every simulation model and narrows down therefore 
the uncertainties of the predictions of all presently available transport models. 
In the light of our findings a thorough investigation of the importance of final state annihilations at all beam energies is in order. At the SPS, data is compatible with considerable anti-particle annihilation \cite{Becattini:2012sq}. At the RHIC the situation is not so clear. Three experiments have published data and show an ambiguous picture. STAR data for example suggests no depletion of the thermal model yield but has not been corrected for weak decays which have shown to be important. Once a meaningful analysis can be done it will be an important step toward a better understanding of the RHIC and LHC physics.

\section{Acknowledgments}
This work has been supported by GSI and Hessian initiative for excellence (LOEWE) 
through the Helmholtz International Center for FAIR (HIC for FAIR), by the European 
Network I3-HP3 Turic and by the agence national de recherche (ANR) program "hadrons@LHC".  J.~S. acknowledges a Feodor
Lynen fellowship of the Alexander von Humboldt foundation. This work was supported by the Office of Nuclear Physics in the US
Department of Energy's Office of Science under Contract No. DE-AC02-05CH11231. The computational resources were provided by the LOEWE Frankfurt Center for Scientific Computing (LOEWE-CSC).

\end{document}